\documentclass[11pt,twoside]{article}
\usepackage[dvips]{rotating}
\usepackage{graphics}
\usepackage{epsfig}
\usepackage{color}
\newenvironment{prcap}[1][16.5cm]{
\hspace*{-0.65cm} \small \begin{minipage}{#1}
\setlength{\parindent}{0.5cm}
\noindent
}
{
\end{minipage}
}

\newcommand{\HA}{\mbox{H$\alpha$}}

\newcommand{\HI}{\mbox{H$\,${\sc i}}}
\newcommand{\HII}{\mbox{H$\,${\sc ii}}}

\definecolor{jmbgray}{gray}{0.3}
\title{{\bf Large-scale star formation in the Magellanic Clouds}}
\author{Jochen~M.~Braun\\
\vspace{0.1cm}\\
\normalsize Sternwarte Univ. Bonn, Auf dem H\"ugel 71, D-53121 Bonn,
 Germany, \\ 
{\tt http://www.astro.uni-bonn.de/$\,\,\tilde{}\,\,$jbraun}\\
}
\date{}
\begin{document}
\setcounter{page}{5}
\maketitle
\def\bull{\vrule height .9ex width .8ex depth -.1ex}
\makeatletter
\def\ps@plain{\let\@mkboth\gobbletwo
\def\@oddhead{}\def\@oddfoot{\hfil\tiny
``Dwarf Galaxies and their Environment'';
Bad Honnef, Germany, 23-27 January 2001; Eds.{} K.S. de Boer, R.-J.Dettmar, U. Klein; Shaker Verlag}%
\def\@evenhead{}\let\@evenfoot\@oddfoot}
\makeatother

\begin{abstract}\noindent
In this contribution I will present the current status of our
project\footnote{The project is a collaborative effort with
Klaas S. de Boer, Martin Altmann, and Holger Schmidt from the Sternwarte Bonn,
Antonella Vallenari from the Astronomical Observatory in Padova,
Ulrich Mebold from the Radioastronomical Institute in Bonn,
\vspace*{0.08cm}
and Jean-Marie Will, formerly at the Sternwarte Bonn, \vspace*{-0.08cm}
in the framework of the Bonn/Bochum - Graduiertenkolleg.}
of stellar population analyses and spatial information of both
Magellanic Clouds (MCs).

The Magellanic Clouds are suitable laboratories and testing ground
for theoretical models of star formation.
With distance moduli of 18.5 and $18.9\;$mag for the LMC and SMC, respectively,
and small galactic extinction, their stellar content can be studied in detail
from the most massive stars of the youngest populations ($< 25\;$Myr)
connected to \HA\ emission
down to the low mass end of about $\frac{1}{10}$ of a solar mass.
Especially the LMC with its large size and small depth ($< 300\;$pc) is
a prefered target to constrain star formation mechanisms.

Based on broad-band photometry ($U,B,V$) I present results for the
supergiant shell (SGS) SMC$\,$1, some regions at the LMC east side incl.
LMC$\,$2 showing different overlapping young populations and the region
around N$\,$171 with its large and varying colour excess, and LMC$\,$4.
This best studied SGS shows a coeval population aged about $12\;$Myr with
little age spread and no correlation to distance from LMC$\,$4's centre.
I will show that the available data are not compatible with many of
the proposed scenarios like SSPSF or a central trigger (like a cluster or GRB),
while a large-scale trigger like the bow-shock of the rotating LMC can do
the job.
\end{abstract}

\vspace*{-13.0cm}

\noindent
{\color{jmbgray}{%
{\footnotesize \it Please note that this verion has low resolution images;
the contribution is available in full quality at the URL:}
'{\footnotesize \tt ftp://ftp.astro.uni-bonn.de/pub/jbraun/gkic/jmb\_dge.ps.gz}'!}}
\vspace*{11.7cm}

\section{Introduction}

On deep pictures of the Magellanic Clouds (MCs) taken
in \HA\ light, Meaburn and collaborators (cf. Meaburn 1980) recognized ten
huge `bubbles' with diameters of 600$\,$--$\,1\,400\;$pc,
the so-called supergiant shells (SGSs or supershells).
These huge shells, according to Goudis \& Meaburn (1978), build a new
group different from the smaller ($< 260\;$pc) giant shells (GSs
or superbubbles).

These structures need very effective creation mechanisms and collisions
of high velocity clouds (HVCs) with the disk of the galaxy or stochastic
self-propagating star formation (SSPSF) have been proposed.
Both might explain the ring of \HII\ regions and the 'hole' in the \HI\ layer,
as observed in the case of LMC$\,$4.
According to the SSPSF model, star formation would `eat'
its way from the initial point to all directions through
the interstellar medium,
creating a big cavity and a thick outer shell of neutral hydrogen ionized
at the inner edge by the early-type stars (O-B2).
Thus one should see a clear age gradient from the centre to the rim.

This contribution will give first results for the SGS SMC$\,$1,
summarize the study at the east side of the LMC with a focus on LMC$\,$2 and
N$\,$171, and ends with a short update (cf. Braun 1998; Braun et al. 1997)
of knowledge about the stellar content of LMC$\,$4 and its implications
on the creation mechanism. 
\vspace*{0.50cm}

\section{First results of a study inside SMC$\,$1}

SMC$\,$1 ($\equiv$ DEM$\,$S$\,$167) is the only supergiant shell (SGS)
detected by Meaburn and collaborators (see e.g. Meaburn 1980, Fig.~3)
in the Small Magellanic Cloud.
It has a diameter of about 600~pc, as indicated by the rough dashed boundaries
in Fig.~1.
Its \HA\ appearance is not as regular as that of LMC$\,$4,
with pronounced emission from south (incl. N$\,90$ $\equiv$~DEM$\,$S$\,$166)
to east, in the northwest, and with a double `rim' in the north.
From the east rim the emission extends toward NE via
N$\,$89 ($\equiv$~DEM$\,$S$\,$164) to N$\,$88 ($\equiv$~DEM$\,$S$\,$161).

The presented photometry has been taken at the $1.54\;$m Danish Telescope at
ESO observatory on Cerro La Silla with  DFOSC and the 2k$^2\;$pix$^2$ LORAL CCD
(W7 Chip; $12.9' \times 12.9'$ field of view)
by me in January 1998 (C,S), and by Martin Altmann in November 1999 (N,W;
only $B,V$).

The CMDs with fitted isochrones in Fig.~2 reveal a small young population of
$\sim 8\;$Myr and a second population of at least $20\;$Myr
(see Braun 2001; Braun et al. 2001b for more details).
\vspace*{0.6cm}

\noindent
\resizebox{8.4cm}{!}{\hspace*{-0.00cm}\includegraphics{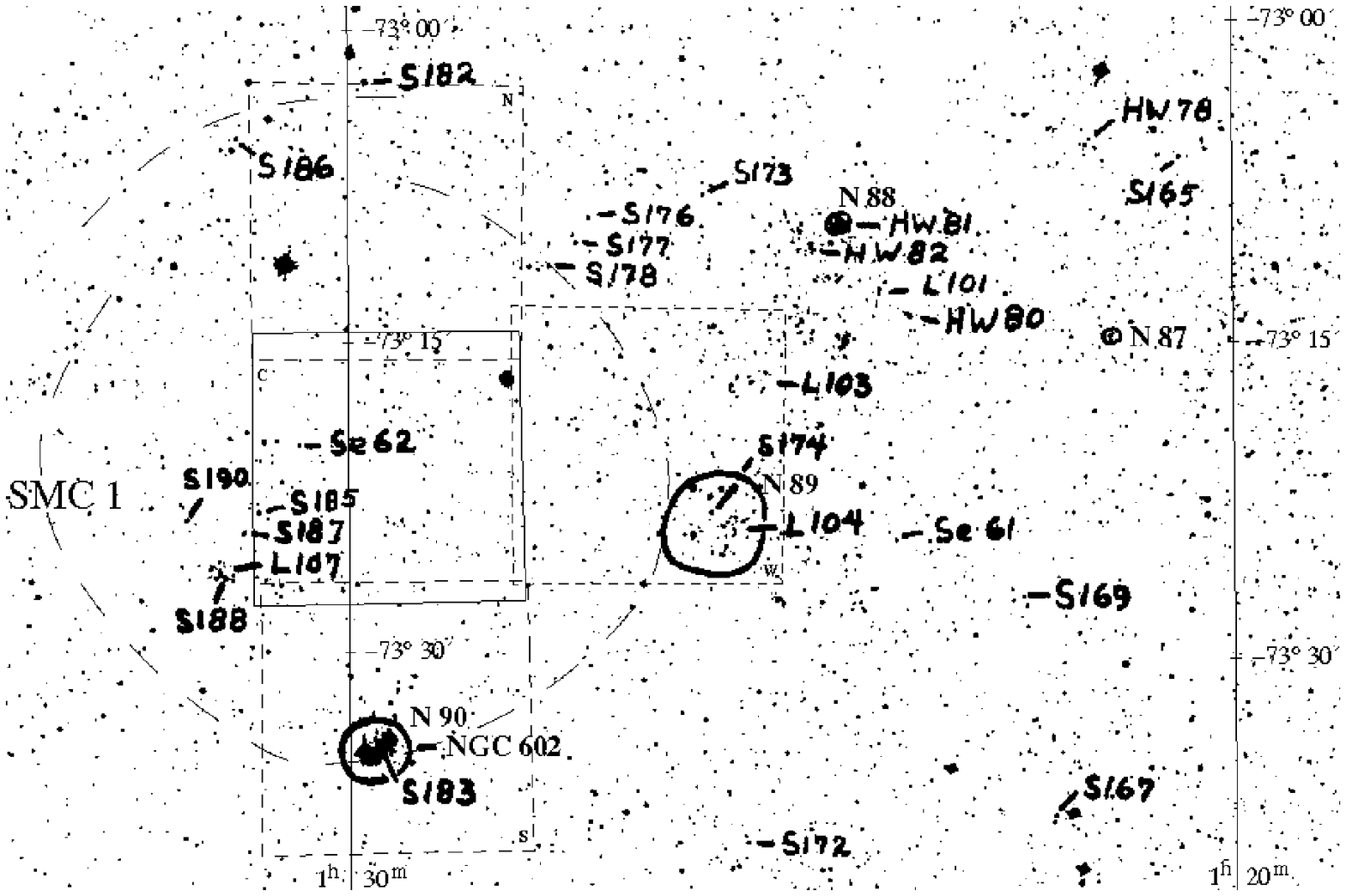}}
\hspace*{0.55cm}
\begin{prcap}[7.7cm]{\bf Fig. 1.}
Mosaic of the central SMC wing region (Shapley 1940) out of $V$ charts
of the Hodge \& Wright (1977) atlas.
The four fields (north, centre, south, west)
of the presented dataset in this supergiant shell are outlined,
with the \HA\ rim of SMC$\,$1 roughly indicated by the dashes.
\vspace*{8.55cm}
\end{prcap} \vspace*{-5.45cm}
%
\vspace*{0.7cm}

\noindent
\epsfig{file=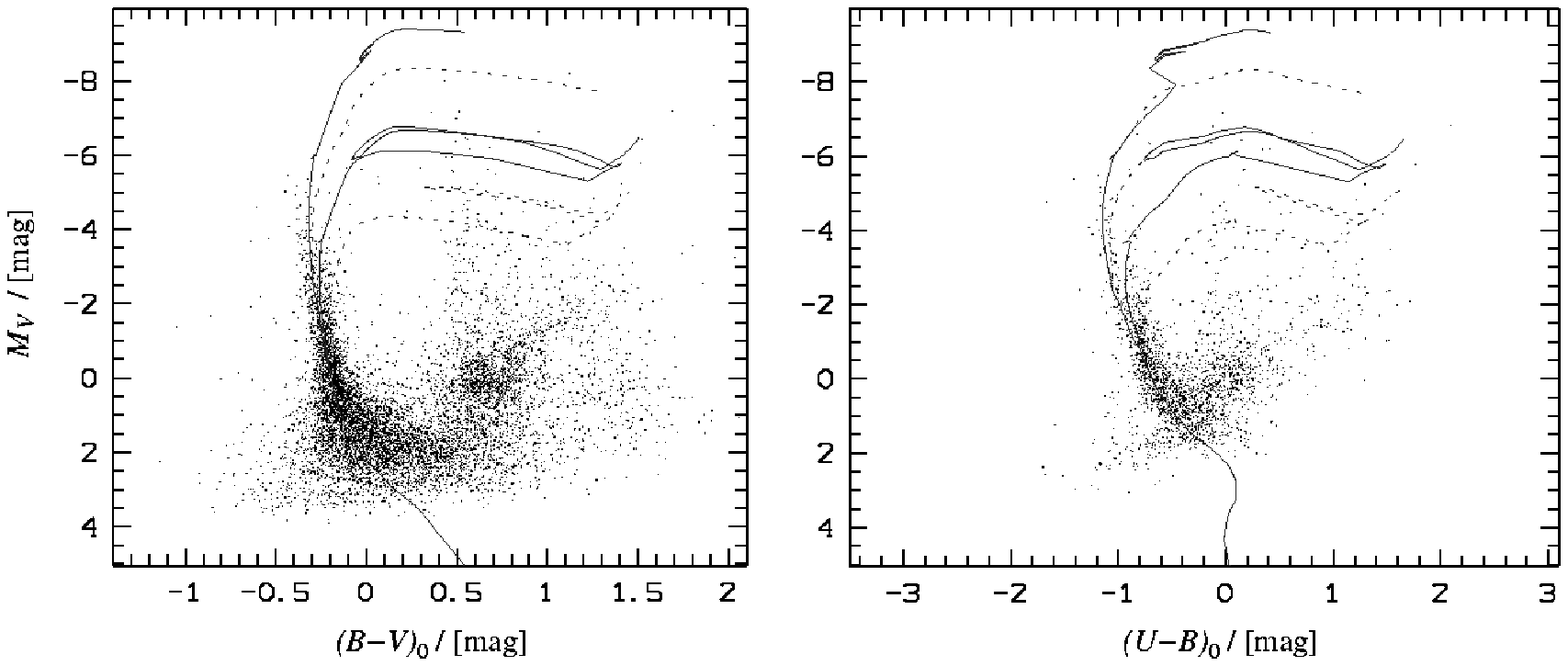,width=16.5cm,angle=0,
        bbllx=56, bblly=316, bburx=544, bbury=522, clip=}
\vspace*{-0.25cm}

\begin{prcap}{\bf Fig. 2.}
CMDs of the entire SMC$\,$1 region (see Fig.~1)
with the isochrones of the Geneva group (Charbonnel et al. 1993)
for SMC metallicity ($Z = 0.004$) of logarithmic ages
$\;\log\,(t/[{\rm yr}]) \in \{ 6.8, 7.0, 7.3, 7.7 \}$ and applied
colour excess of $\,E_{B-V} = 0.12-0.19\;$mag.
The $B-V$ diagram contains $8\,704$ data points,
the $U-B$ a total of $2\,166$.
\end{prcap}
\newpage

\section{Stellar populations at the LMC east side}

In 1996 I observed five regions located at the LMC east side with
10~CCD~positions in $U,B,V$ passbands with the $1.54\,$m Danish Telescope.
These regions (see Fig.~3) are from north to south:
the giant shell N$\,$70 (region A$_{1-2}$, cf. Braun 1998),
a strip inside SGS LMC$\,$2 (B$_{1-4}$),
an outer field east of LMC$\,$2 (C),
the region around N$\,$171 (D),
and the southern part of N$\,$214 (E$_{1-2}$).
Here I will concentrate on LMC$\,$2 and N$\,$171 data
(see Braun 2001; Braun et al. 2001a for further information).

\vspace*{0.35cm}

\noindent
\epsfig{file=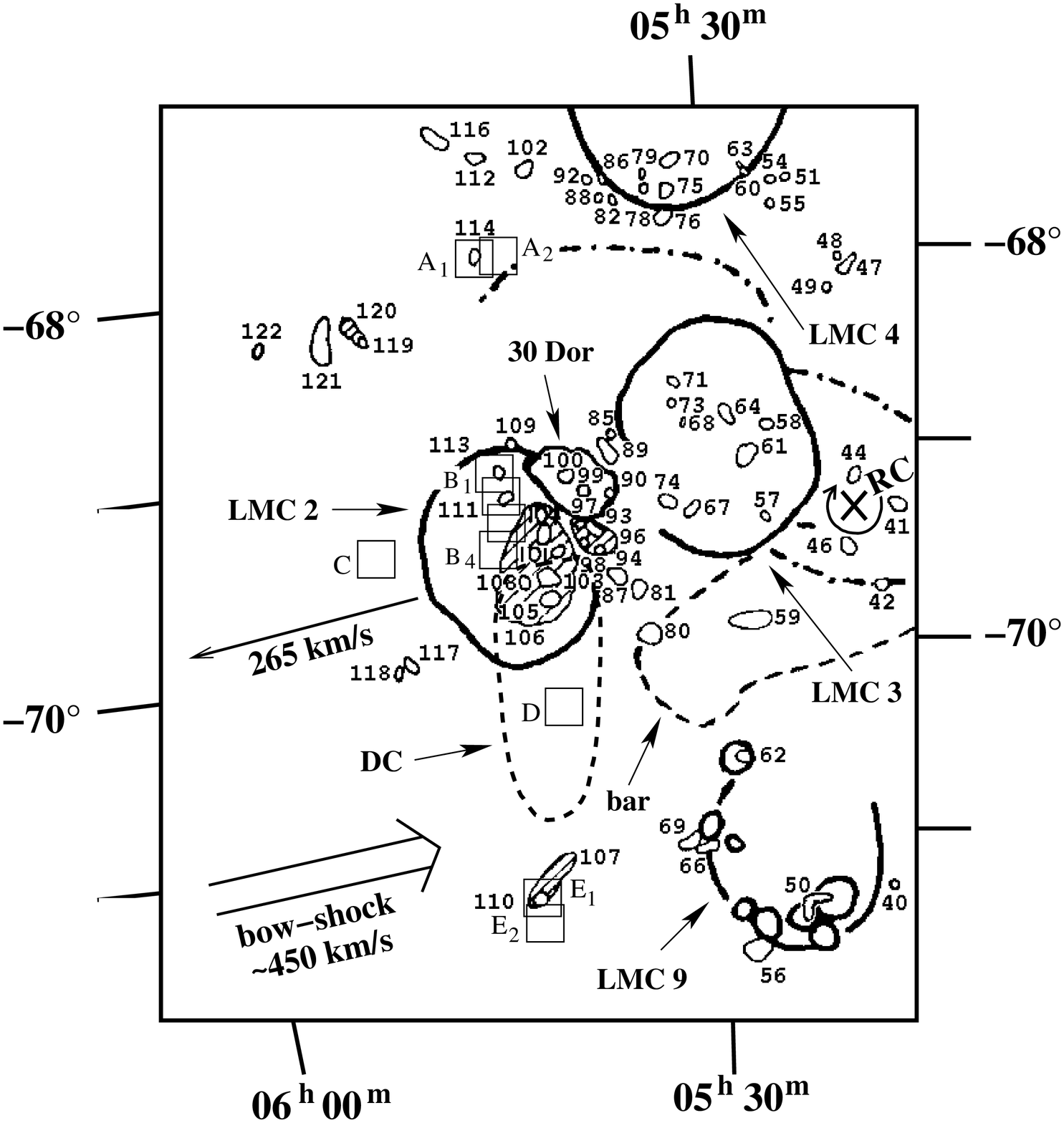,width=8.0cm,angle=0,
        bbllx=12.0, bblly=25.0, bburx=760.0, bbury=820.0, clip=}
\hspace*{0.65cm}
\begin{prcap}[8.0cm]
\vspace*{-11.3cm}

\noindent
{\bf Fig.~3.}
Sketch of the large-scale features (see Mizuno et al. 2001 for a detailed
map of the dark cloud, DC)
and OB associations at the LMC east side.
Additionally, the 5 regions (A-E) out of 10 CCD positions are indicated.
The position of the rotation centre, RC, and
the movement of the LMC in the galactic halo are indicated by
the direction of rotation and proper motion.
The resulting bow-shock, driven by the sum of motion and rotation, is
$\sim 450\;$km s$^{-1}$ (see de Boer et al. 1998; de Boer 1998).
\end{prcap}\vspace*{0.25cm}

%
\noindent
\epsfig{file=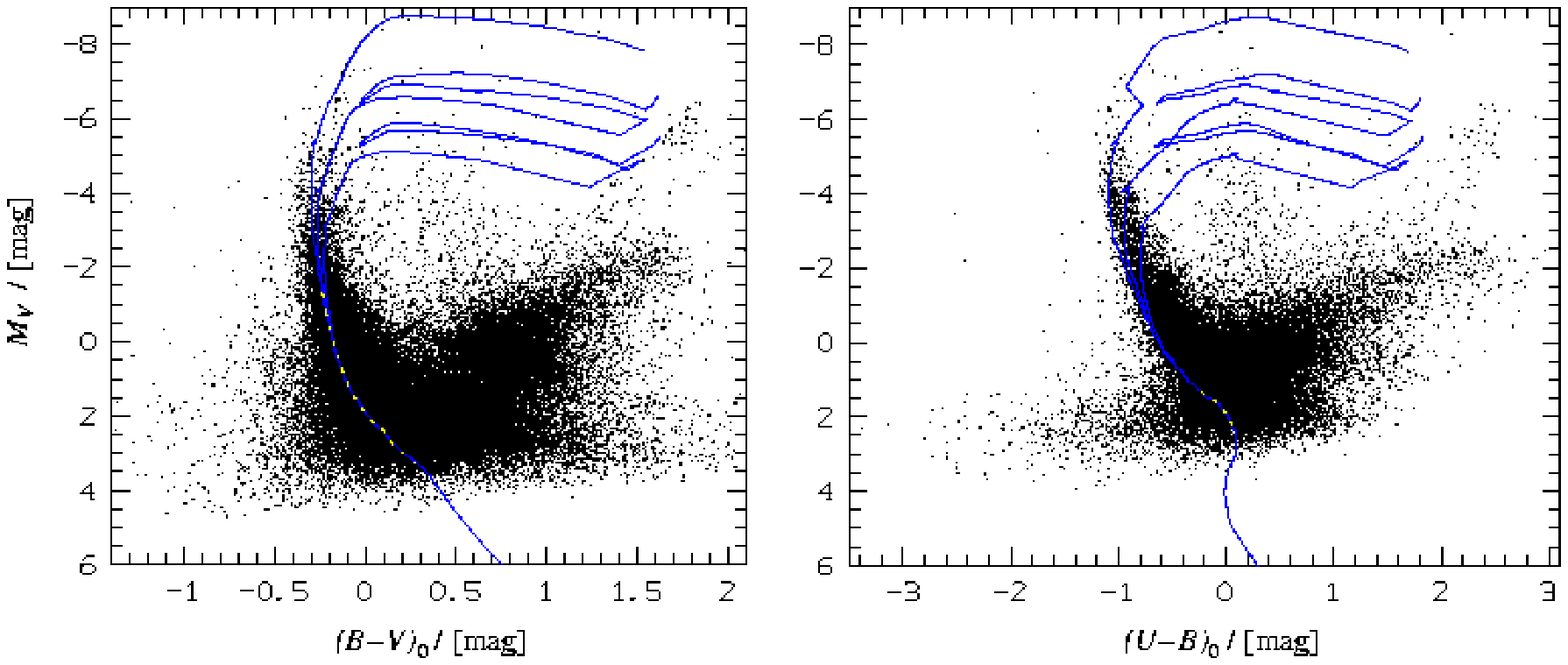,width=11.45cm,angle=0,
        bbllx=42, bblly=313, bburx=554, bbury=529, clip=}
\hspace*{0.45cm}
\begin{prcap}[4.65cm]
\vspace*{-6.40cm}

\noindent
{\bf Fig.~4.}
CMDs of the entire strip inside LMC$\,$2 (B$_{1-4}$)
with Geneva isochrones (see~Schaerer et al. 1993; $Z = 0.008$) of
logarithmic ages $\;\log\,(t/[{\rm yr}]) \in \{ 6.9, 7.2, 7.5 \}$ and
applied \mbox{$V$ extinction} of 0.68$\;$mag.
\end{prcap} \vspace*{0.00cm}

\vspace*{0.15cm}

\noindent
\epsfig{file=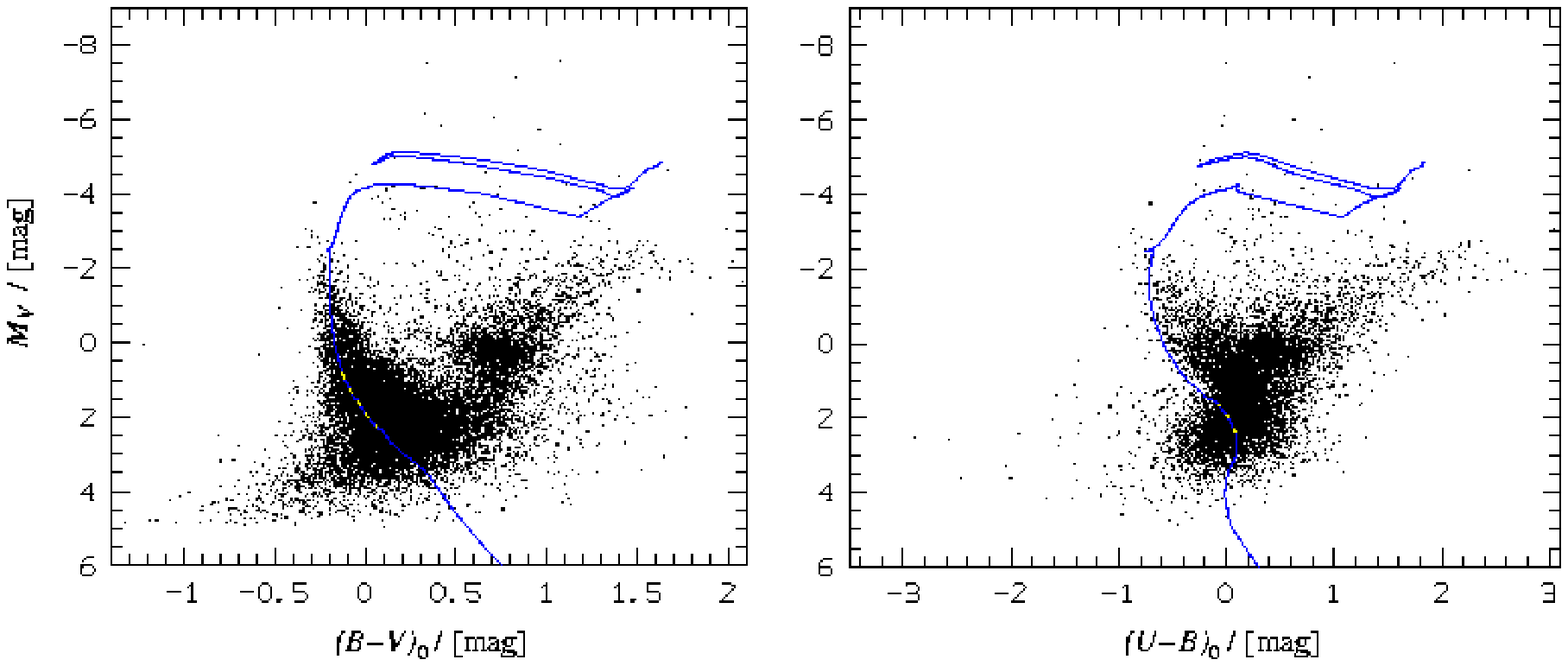,width=11.45cm,angle=0,
        bbllx=37, bblly=312, bburx=557, bbury=531, clip=}
\hspace*{0.45cm}
\begin{prcap}[4.65cm]
\vspace*{-6.70cm}

\noindent
{\bf Fig.~5.}
CMDs of the~outer field east of LMC$\,$2 (C)
with Geneva isochrones (Schaerer et al. 1993; $Z = 0.008$) of 
logarithmic ages $\;\log\,(t/[{\rm yr}]) = 7.7$ and 
applied \mbox{$V$ extinction} of 0.65$\;$mag.
\end{prcap}
\newpage

\vspace*{0.45cm}

\noindent
\epsfig{file=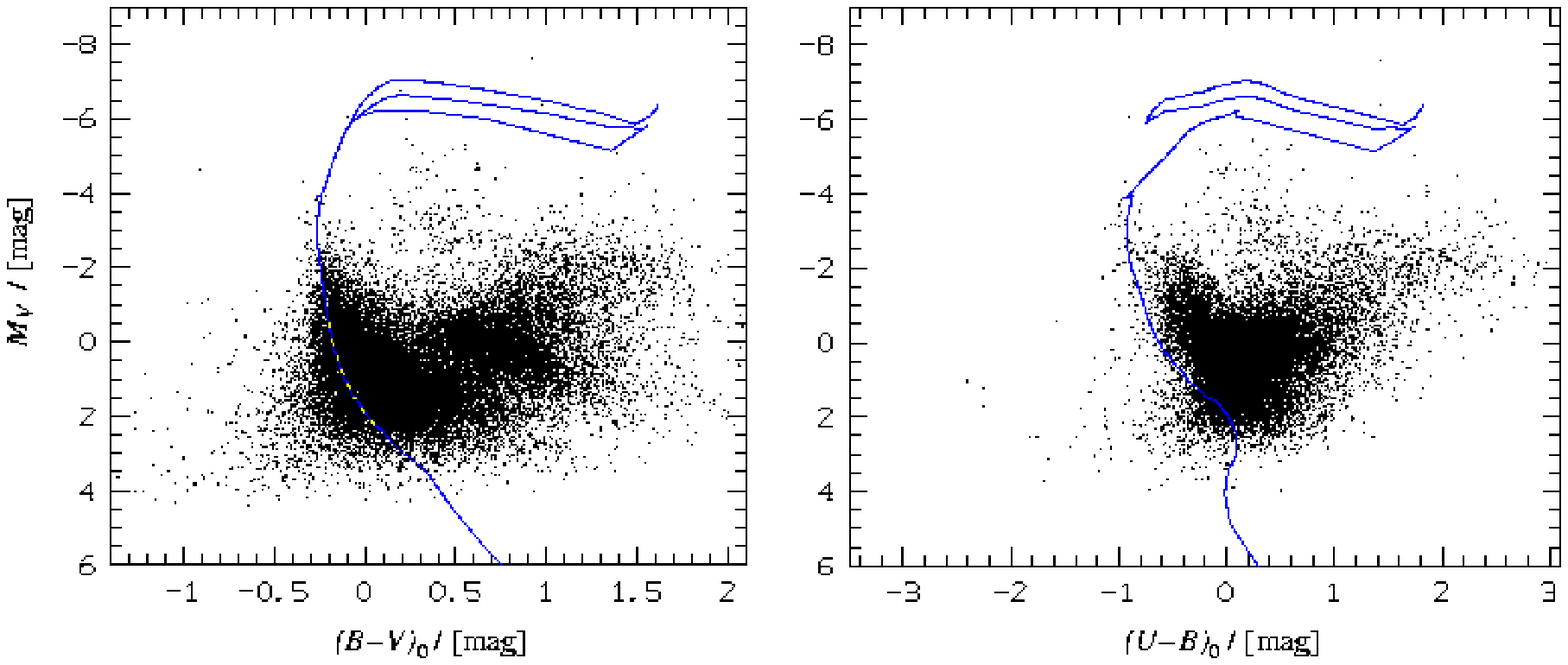,width=16.3cm,angle=0,
        bbllx=36, bblly=310, bburx=556, bbury=529, clip=}
\vspace*{0.25cm}

\begin{prcap}{\bf Fig.~6.}
CMDs of the region around N$\,$171 (D)
with Geneva isochrones (Schaerer et al. 1993; $Z = 0.008$) of
logarithmic ages $\;\log\,(t/[{\rm yr}]) = 7.25$ and
applied \mbox{$V$ extinction} of 0.96$\;$mag.
One should note that the isochrone has only been fitted to the youngest and
less reddened stars of the $B-V$ diagram.
\end{prcap} \vspace*{0.25cm}

Figs.~4 and 5 show the CMDs of the entire strip inside LMC$\,$2 and of the
outer field, respectively.
While the CMDs of the outer field only show a stellar population of about
$50\;$Myr and older,
the CMDs of the strip containing N$\,$164, NGC$\,$2100, and NGC$\,$2102
reveal a mixture of young populations (present in all four fields), covering
the interval from 8 to 32~Myr.
While the two southern fields are similar, the two northern show differences
in the age structure by a component younger than 8~Myr concentrated on
N$\,$164 (B$_1$) and a 16~Myr component of NGC$\,$2100.

For N$\,$171, the youngest population discovered in the CMDs (Fig.~6) has
an age of about 18~Myr.
This region was selected at the locus of an X-ray shadowing region
(cf. Blondiau et al. 1997) and shows extraordinary reddening with a large
spread.
The colour extinctions ($E_{B-V}$/[mag]) derived from the reddening free
parameter Q (peak of the distribution with FWHM in '($\,$)') and from
the isochrone fit (in '[~]') are:
0.075 (0.08) [0.11] for N$\,$70 (A$_1$),
0.14  (0.08) [0.21] for the outer field east of LMC$\,$2,
0.17  (0.11) [0.19] for N$\,$214,
0.185 (0.16) [0.22] for the LMC$\,$2 strip, and
{\bf 0.29  (0.24) [0.31]} for the {\bf N$\,$171} region.
One should note that the peak of the distribution is shifted from the mean and
that the isochrone fit is based on the youngest and for N$\,$214 not numerous
population,
which may have a different reddening than most other stars (maybe due to
unrevealed 3d structure).

\section{Supergiant shell LMC$\,$4 and its large-scale triggered origin}

Our first $B,V$ photometry inside LMC$\,$4 (a 'J'-shaped region, see Fig.~7)
taken in 1993 with the $0.91\;$m Dutch Telescope (Braun et al. 1997; Braun 1998)
has shown, as given in Fig.~8b, that the central superassociation, LH$\,$77,
has an age of $11-13\;$Myr.
This population is present in the entire interior of this supergiant shell
and at the inner rim region (e.g. LH$\,$63).
The associations at the rim tend to show age gradients with younger ages for
associations more distant from the inner rim of \HA\ filaments (see
Gouliermis et al. 2001; Table~6 of Braun et al. 1997).

Thus, this huge coeval population gave rise to the need for a large-scale
trigger, like an infalling cloud or a large shock front created by the movement
of the LMC through the galactic halo (de~Boer et al. 1998; de~Boer 1998),
while models with a central trigger or a propagation could be ruled out.
SSPSF is therefore only valid on small scales (up to $\sim 300\;$pc) as it
is visible on the rim, while the creation mechanism of LMC$\,$4 has its origins
on larger ($\sim 1\;$kpc) scales.
\newpage

\noindent
\epsfig{file=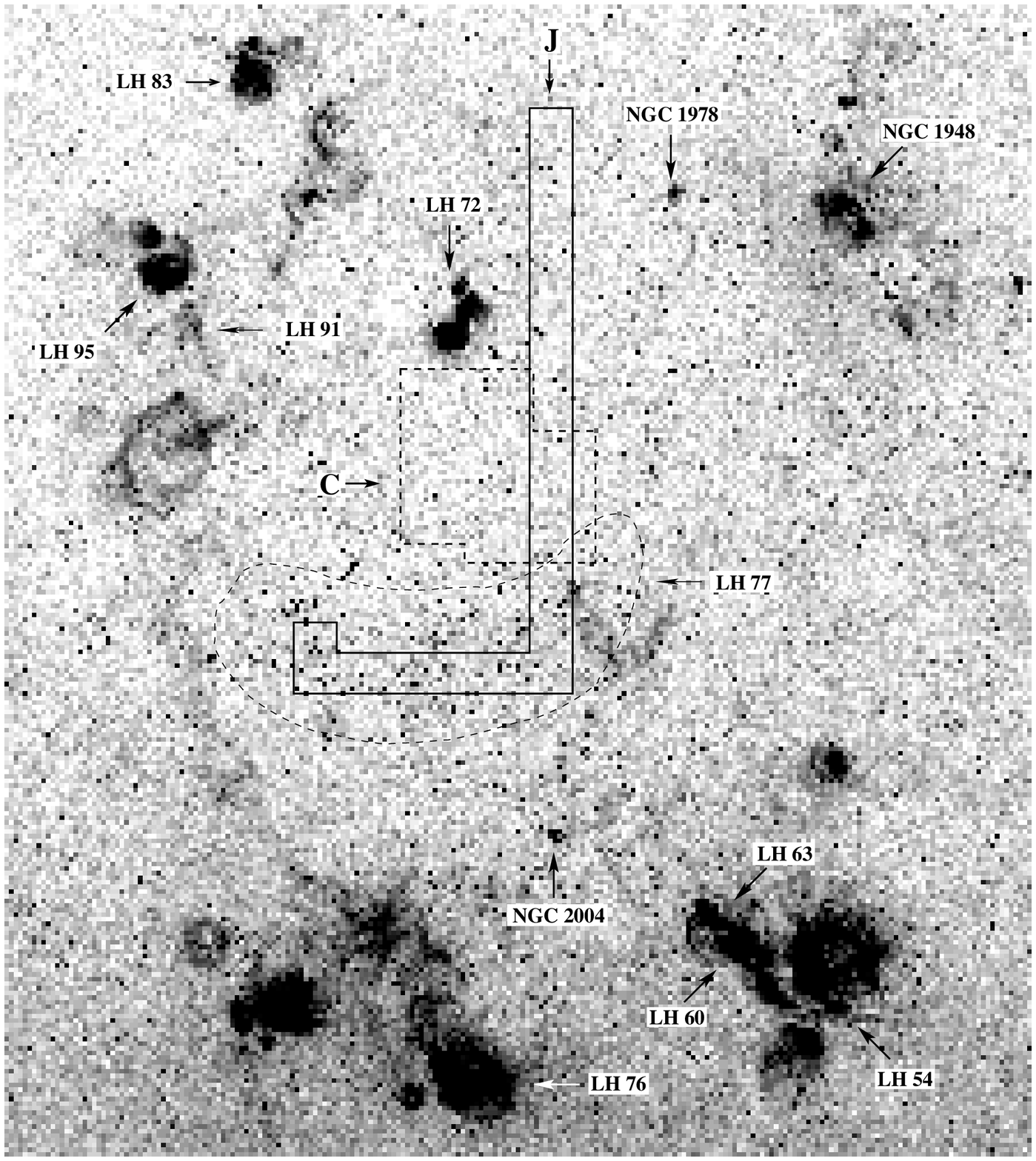,width=8.5cm,angle=0,
        bbllx=1, bblly=2, bburx=537, bbury=602, clip=}
\hspace*{0.15cm}
\epsfig{file=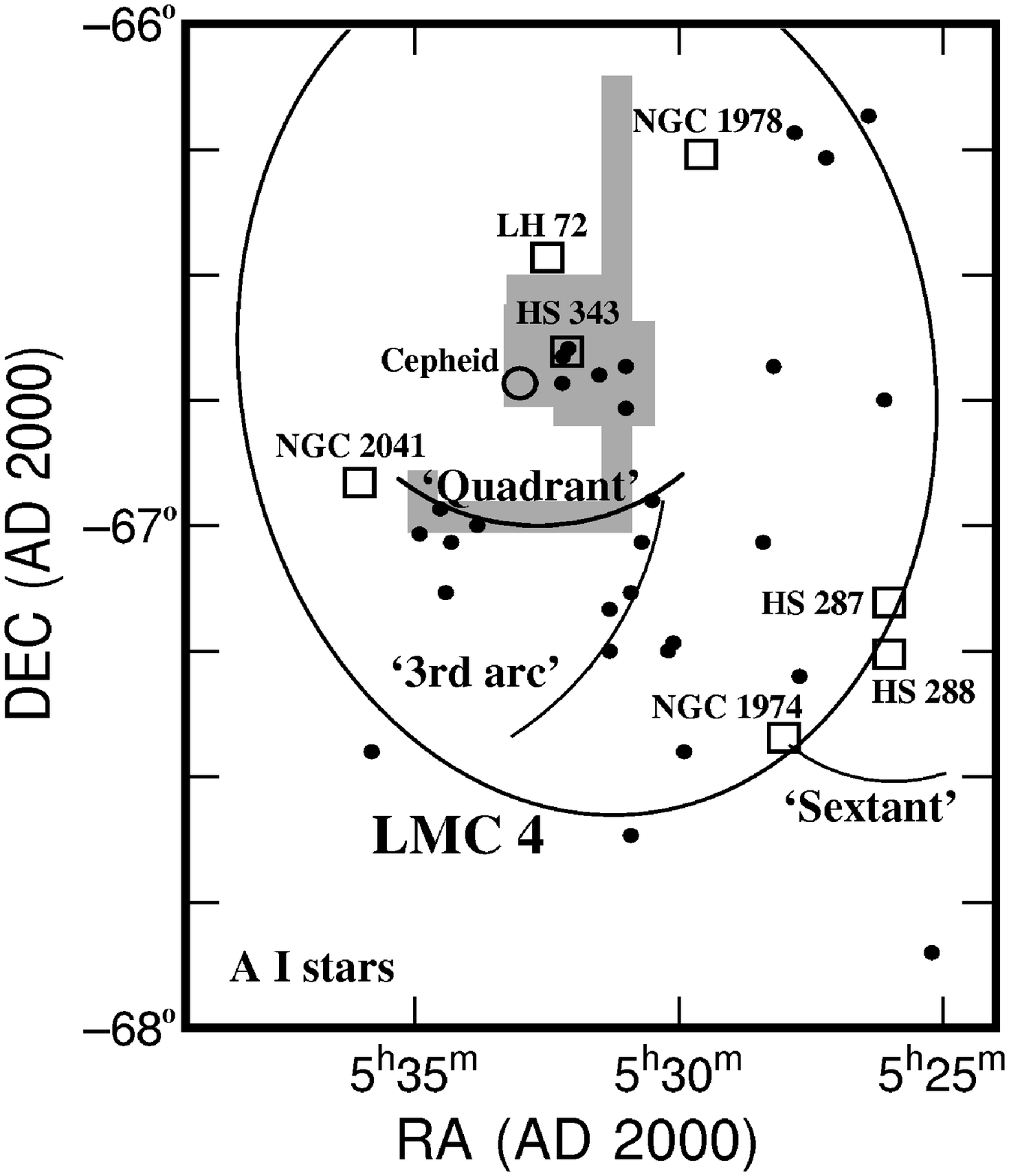,width=7.65cm,angle=0,
        bbllx=40, bblly=120, bburx=558, bbury=722, clip=}
\vspace*{-0.10cm}

\begin{prcap}{\bf Fig. 7.}
a) \HA\ image of supergiant shell LMC$\,$4 [left panel] made from a scan of
a photographic plate taken with the Curtis Schmidt Telescope
at Cerro Tololo (Kennicutt \& Hodge 1986).
The locations of some stellar associations and the borders of the two
datasets (J and C, respectively) are marked.
b)~Sketch~of selected structures [right panel] near LMC$\,$4 (adapted
from Fig.~2 of Efremov~\& Elmegreen 1998).
\end{prcap}\vspace*{0.35cm}
%
\vspace*{0.45cm}

\noindent
\epsfig{file=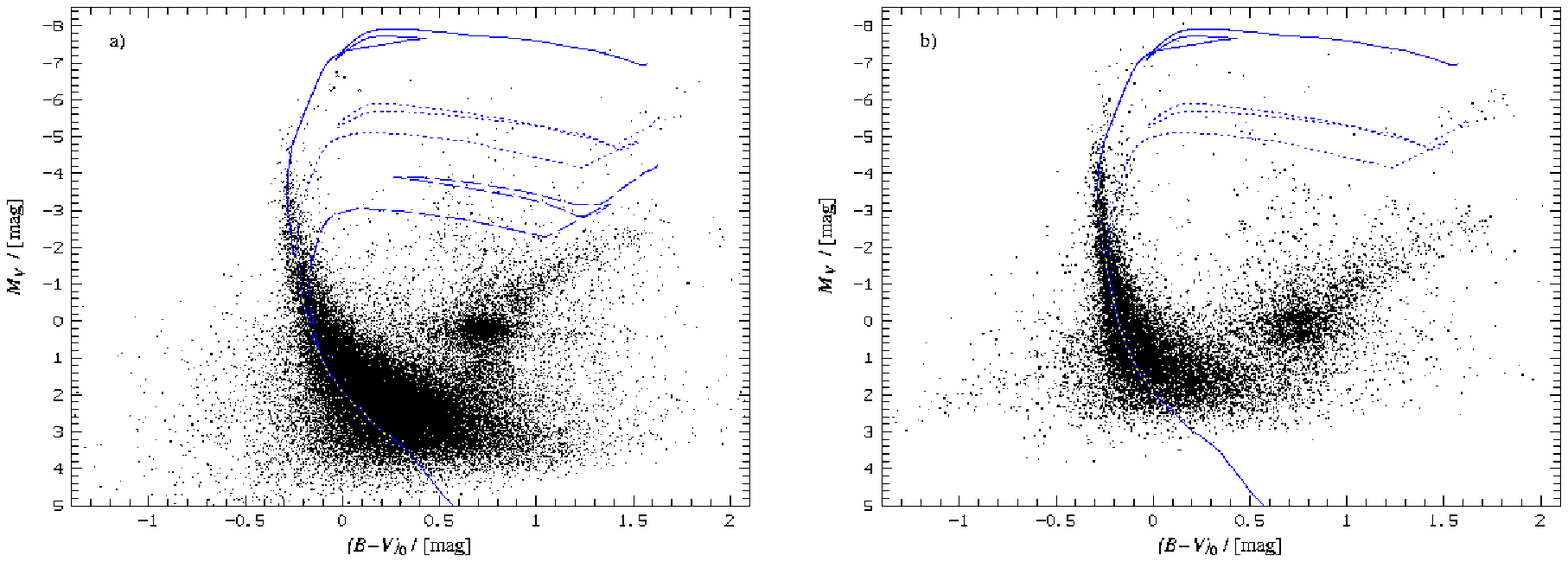,width=16.5cm,angle=0,
        bbllx=38, bblly=330, bburx=557, bbury=513, clip=}
\vspace*{-0.15cm}

\begin{prcap}{\bf Fig. 8.}
CMDs of LMC$\,$4 with Geneva isochrones (Schaerer et al. 1993; $Z = 0.008$) of
logarithmic ages $\;\log\,(t/[{\rm yr}]) \in \{ 7.05, 7.5, 8.0 \}$ and
extinction correction of $A_V = 0.31\;$mag.\\
a) CMD with $46\,749$ data points of the central data set (C, see Fig.~7a)
of LMC$\,$4 [left panel].\\
b) CMD with $15\,787$ d. p. of the LMC$\,$4 'J' dataset (Braun et al.
   1997) without fields 11-16 [right panel].\\
In panel a) the crosses near $M_V \approx -6.5\;$mag are six A~supergiants also
marked in Fig.~7b.
\end{prcap} \vspace*{0.25cm}

Nevertheless, after these results further models with central trigger had been
proposed by Efremov \& Elmegreen (1998), describing three stellar arcs as
triggered by central clusters (as given by the bows and boxes in Fig.~7b).
Additionally, Efremov \& Fargion (2000) presented a scenario,
by which cluster arcs (now even four in the LMC$\,$4 region and vicinity)
are each pressurized by the collimated and precessing beam of a $\gamma$-ray
burst / soft gamma repeater (originating in NGC$\,$1978).
\newpage

To further constrain the triggering mechanism, we observed
three central fields (C) of LMC$\,$4 (see Fig.~7a and Braun et al. 2000)
in $B$ and $V$.
The resulting colour-magnitude diagram (Fig.~8a) has the same morphology as
the one of the 'J' dataset excluding the overlapping region~(Fig.~8b).
The two clusters (HS$\,$343 and KMHK$\,$1000), which may have triggered
star formation after Efremov \& Elmegreen, are much too old
(0.1~Gyr and 0.3~Gyr, respectively) to have stimulated further star formation
in the birth cloud of LH$\,$77.
The triggering of different arcs does neither fit to a homogeneous coeval
population nor to the morphology of LMC$\,$4 in \HI\ and \HA.

With the stringent results of the good-quality\footnote{The comparison of
 the three photometries (J, C, and Dolphin \& Hunter 1998 [AJ 116, 1275])
 shows good agreement.
 All three datasets are available electronically,
 see CDS {\tt J/A+A/328/167}, {\tt J/AJ/116/1275}, and AIUB FTP account,
 linked from the WWW pages given above (Braun et al. 2000).}
photometry available,
it seems to be inevitable that LMC$\,$4 has been driven and ionized by the
massive stars of the coeval population inside the SGS, with about
5000 stars already exploded in supernovae of type II.
Thus, the gas had been driven out of the central region in a turbulent fashion,
causing an outbreak at the rear side and a component approaching with
11~km/s (Domg\"orgen et al. 1995).
The star burst 12~Myr ago has been most probably driven by the bow-shock,
resulting in a compression zone. Its location is currently, due to the LMC
rotation (see Fig.~3), at the south eastern edge and thus may have caused
the CO complex south of 30$\,$Doradus (see Mizuno et al. 2001).

The updated and enhanced content of the articles
(Braun et al. 1997, 2000, 2001a,b),
incl. first results on SGSs LMC$\,$7 and LMC$\,$1,
can be found with full discussion in Braun (2001).
\vspace*{0.15cm}

{\small
\begin{description}{} \itemsep=0pt \parsep=0pt \parskip=0pt \labelsep=0pt
\item {\bf References}\vspace*{0.15cm}

\item Blondiau, M.J., Kerp, J., Mebold, U., Klein, U. 1997,
A\&A 323, 585

\item Braun, J.M. 1998,
``The~Magellanic~Clouds and Other Dwarf~Galaxies'', eds T.~Richtler,
J.M.~Braun, Shaker~Verlag, Aachen, ISBN~3-8265-4457-9, pp.~115-120 \\
(see {\tt http://www.astro.uni-bonn.de/$\,\,\tilde{}\,\,$webstw/ws98/jmb\_pt.html})

\item Braun, J.M. 2001, Ph.~D. Thesis, University of Bonn,
Shaker Verlag, Aachen, ISBN~3-8265-8325-6\\
({\footnotesize expected for 11/2001,}
see {\tt http://www.astro.uni-bonn.de/$\,\,\tilde{}\,\,$jbraun/phdt.html})

\item Braun, J.M., Bomans, D.J., Will, J.-M., de Boer, K.S. 1997, A\&A 328, 167
({\sf arXiv:astro-ph/9708081})

\item Braun, J.M., de Boer, K.S., Altmann, M. 2000,
{\sf arXiv:astro-ph/0006060}\\ (see
 {\tt http://www.astro.uni-bonn.de/$\,\,\tilde{}\,\,$jbraun/phdt\_lmc4c.html})

\item Braun, J.M., Vallenari, A., de Boer, K.S. 2001a, in prep.\\
(see {\tt http://www.astro.uni-bonn.de/$\,\,\tilde{}\,\,$jbraun/phdt\_lmce.html})

\item Braun, J.M., Altmann, M., de Boer, K.S. 2001b, in prep.\\
(see {\tt http://www.astro.uni-bonn.de/$\,\,\tilde{}\,\,$jbraun/phdt\_smc1.html})

\item Charbonnel, C., Meynet, G., Maeder, A., Schaerer, D. 1993,
A\&A 101, 415

\item de Boer, K.S. 1998,
``The~Magellanic~Clouds and Other Dwarf~Galaxies'', eds T.~Richtler,
J.M.~Braun, Shaker~Verlag, Aachen, ISBN~3-8265-4457-9, p.~125

\item de Boer, K.S., Braun, J.M., Vallenari, A., Mebold, U. 1998, A\&A 329, L49
({\sf arXiv:astro-ph/9711052})

\item Domg\"orgen, H., Bomans, D.J., de Boer, K.S. 1995, A\&A 296, 523

\item Efremov, Y.N., Elmegreen, B.G. 1998,
MNRAS 299, 643 ({\sf arXiv:astro-ph/9805092})

\item Efremov, Y.N., Fargion, D. 2000,
A\&A subm. ({\sf arXiv:astro-ph/9912562})

\item Goudis, C., Meaburn, J. 1978, A\&A 68, 189

\item Gouliermis, D., et al.
2001, 
``Dwarf Galaxies and their Environment'', eds K.S.~de~Boer, R.J.~Dettmar,
U.~Klein,
Shaker Verlag, Aachen, in press (these proceedings)

\item Hodge, P.W., Wright, F.W. 1977, University of Washington Press,
Seattle, London

\item Kennicutt, R.C., Hodge, P.W. 1986, ApJ 306, 130

\item Meaburn, J. 1980, MNRAS 192, 365

\item Mizuno, N., Yamaguchi, R., Onishi, T., Mizuno, A., Fukui, Y.,
2001, 
``Dwarf Galaxies and their Environment'', eds K.S.~de~Boer, R.J.~Dettmar,
U.~Klein,
Shaker Verlag, Aachen, in press (these proceedings)

\item Schaerer, D., Meynet, G., Maeder, A., Schaller, G. 1993, A\&AS 98, 523

\item Shapley, H. 1940, Harvard Obs. Bull. 914, 8
\end{description}
}
\end{document}